\documentclass[runningheads]{llncs}
\usepackage[T1]{fontenc}
\usepackage{graphicx}
\usepackage{amsmath}
\usepackage{amssymb}
\usepackage{booktabs}

\makeatletter
\def\ps@headings{\let\@mkboth\@gobbletwo
  \let\@oddfoot\@empty\let\@evenfoot\@empty
  \def\@evenhead{\normalfont\small\hspace{\headlineindent}%
                 \leftmark\hfil}
  \def\@oddhead{\normalfont\small\hfil\rightmark\hspace{\headlineindent}}
  \def\chaptermark##1{}%
  \def\sectionmark##1{}%
  \def\subsectionmark##1{}}
\makeatother
\begin{document}
\title{DeepHSIC: Deep Learning-based Signal Detector for Hybrid Downlink IM-NOMA}

\titlerunning{DeepHSIC for Hybrid Downlink IM-NOMA}
%
%
\author{Dung Nguyen Tran\inst{1,5} 
\and
Toan D. Gian\inst{2} 
\and
Tien-Hoa Nguyen\inst{5}
\and
Mai Xuan Trang\inst{3}\thanks{Corresponding author.}
\and
Tien-Cuong Nguyen\inst{4}
\and
Thien Van Luong\inst{1} 
}
\authorrunning{D. N. Tran et al.}
%
\institute{Business AI Lab, College of Technology, National Economics University, Vietnam\\
\and
College of Engineering, Northeastern University, Boston, MA, USA\\
\and
A2I Lab, Phenikaa School of Computing, Phenikaa University, Hanoi, Vietnam\\
\and
VNPT AI, VNPT Group, Hanoi, Vietnam\\
\and
School of EEE, Hanoi University of Science and Technology, Hanoi, Vietnam\\
\email{trandung280502@gmail.com, toan.g@northeastern.edu, hoa.nguyentien@hust.edu.vn, trang.maixuan@phenikaa-uni.edu.vn, nguyentiencuong@vnpt.vn,   thienlv@neu.edu.vn}}
\maketitle              
\begin{abstract}
DeepHSIC is introduced as a neural receiver for hybrid downlink IM-NOMA transmission. The considered scheme combines power-domain NOMA with a composite OFDM/OFDM-IM waveform, so that user information is mapped jointly onto constellation symbols, subcarrier-index patterns, and different power levels. Although maximum-likelihood detection can achieve strong reliability for this model, its search space grows rapidly with the number of users and subcarriers. Conventional SIC reduces part of this burden, but its sequential cancellation may still accumulate errors and does not fully exploit the structure of IM-NOMA signals. To address this limitation, the proposed detector embeds dedicated deep neural network modules into the receiver and replaces the most computationally demanding SIC operations with learned inference blocks. The receiver is trained for Rayleigh fading channels and uses preprocessed channel-output features to recover user symbols. Simulation results show that DeepHSIC reaches BER performance close to model-based detectors under both perfect and imperfect CSI while requiring substantially lower detection time. These results indicate that learned SIC-style detection is a practical candidate for scalable hybrid downlink IM-NOMA receivers.


\end{abstract}
\section{Introduction}
The rapid expansion of user connectivity has placed stronger requirements on wireless systems in terms of efficiency, reliability, and scalability \cite{wang2020}. In response to these requirements, Orthogonal Frequency Division Multiplexing (OFDM) and its related variants have become important transmission techniques for modern wireless communications. OFDM is widely adopted because it can effectively cope with frequency-selective fading and multipath propagation while maintaining high spectral utilization. Nevertheless, the continuous growth in traffic volume, data-rate demand, and spectrum scarcity has encouraged the development of more advanced OFDM-based schemes. One promising direction is to combine conventional OFDM with the principle of Index Modulation (IM) \cite{IMfor5G}, leading to OFDM with Index Modulation (OFDM-IM) \cite{OFDM-IM}. In this technique, information is carried not only by the constellation symbols, such as their amplitude and phase, but also by the activation pattern of selected subcarriers.


By allowing only a portion of subcarriers to be active during transmission, OFDM-IM can improve spectrum usage while also lowering the peak-to-average power ratio compared with conventional OFDM systems  \cite{M-MAfor5G}. The indices of the active subcarriers are exploited as an additional information-bearing domain, which enables extra bits to be transmitted without requiring all subcarriers to carry modulated symbols. Owing to this mechanism, OFDM-IM is considered a suitable alternative in communication scenarios where spectral resources are constrained. Its main benefit is the ability to provide a practical compromise among spectral efficiency, energy consumption, and implementation complexity, making it a potential candidate for future wireless network designs \cite{Basar2013}.
 

In addition to modulation techniques, multiple access schemes are essential for supporting the massive connectivity expected in next-generation networks. Among these schemes, Non-Orthogonal Multiple Access (NOMA) has attracted considerable attention because it can increase system capacity and improve user fairness by enabling different users to occupy the same time-frequency resources \cite{Ding2017}. In power-domain NOMA, the transmitter superimposes user signals with different power allocations, whereas the receiver applies successive interference cancellation (SIC) to recover the individual signals from the composite received waveform. However, SIC performs effectively only when the received power levels of the users are sufficiently distinguishable. For this reason, OFDM and OFDM-IM have been explored to support the coexistence of grant-free ultra-reliable low-latency communication (URLLC) and grant-based latency-tolerant services \cite{uplink-IM-NOMA}. When URLLC traffic appears, the available resources can be shared between these services, while K-repetition transmission and maximum-ratio combining are employed to strengthen URLLC reliability. Furthermore, the NOMA principle has also been combined with space-domain IM in \cite{uplink-IM-NOMA}, \cite{Downlink-IM-NOMA}.


In the current work, we investigate the Hybrid IM-NOMA system \cite{hybrid}, which integrates conventional OFDM and OFDM-IM through the superposition coding mechanism of NOMA.
In this scheme, subcarriers are selectively activated based on index information, and distinct power levels are assigned to users within the same frequency band. This enables the concurrent transmission of multiple data streams with minimal mutual interference, as each stream is distinguished by its power level and subcarrier index. This system offers numerous benefits, including enhanced spectral efficiency by multiplexing multiple data streams over the same spectrum and improved energy efficiency through selective subcarrier activation \cite{Basar2016}. The flexibility in adapting power levels and subcarrier usage according to user conditions allows for robust and efficient communication, accommodating varying channel conditions and user requirements. This adaptability is particularly beneficial in diverse communication environments where user conditions can vary significantly \cite{Mao2018}. However, implementing Hybrid IM-NOMA systems presents significant challenges, especially in the area of signal detection. Traditional signal detection techniques such as maximum likelihood (ML) detection, although optimal, are computationally prohibitive when dealing with a large number of users and subcarriers. Meanwhile, SIC more computationally feasible, suffers from issues like error floor \cite{uplink-IM-NOMA} and is inefficient when the power levels of users are not sufficiently distinct. These challenges emphasis the need for more advanced and efficient signal detection approach.

Leveraging the power of deep learning (DL), various studies \cite{Toan2022APSIPA,SICnet,Toan2023,Thien2019} have successfully employed DL for signal detection in wireless networks. Inspired by this, we introduce a novel deep learning-based detector, termed DeepHSIC, specifically designed for hybrid IM-NOMA systems. DeepHSIC replaces the computational block in the SIC detector with dedicated deep learning blocks, leveraging the power of deep neural networks (DNNs) to perform signal detection \cite{van2022deep}. This innovative approach offers near-optimal performance compared to traditional ML and SIC techniques, while maintaining low complexity. The DeepHSIC detector is capable of handling the complexities of the hybrid modulation scheme, providing a robust solution that scales well with system size and complexity. 
Through this learning capability, the proposed detector can identify hidden patterns in noisy received signals, thereby improving detection reliability while reducing the computational cost required for practical implementation.

\begin{figure}
\begin{centering}
\includegraphics[width=0.8\textwidth]{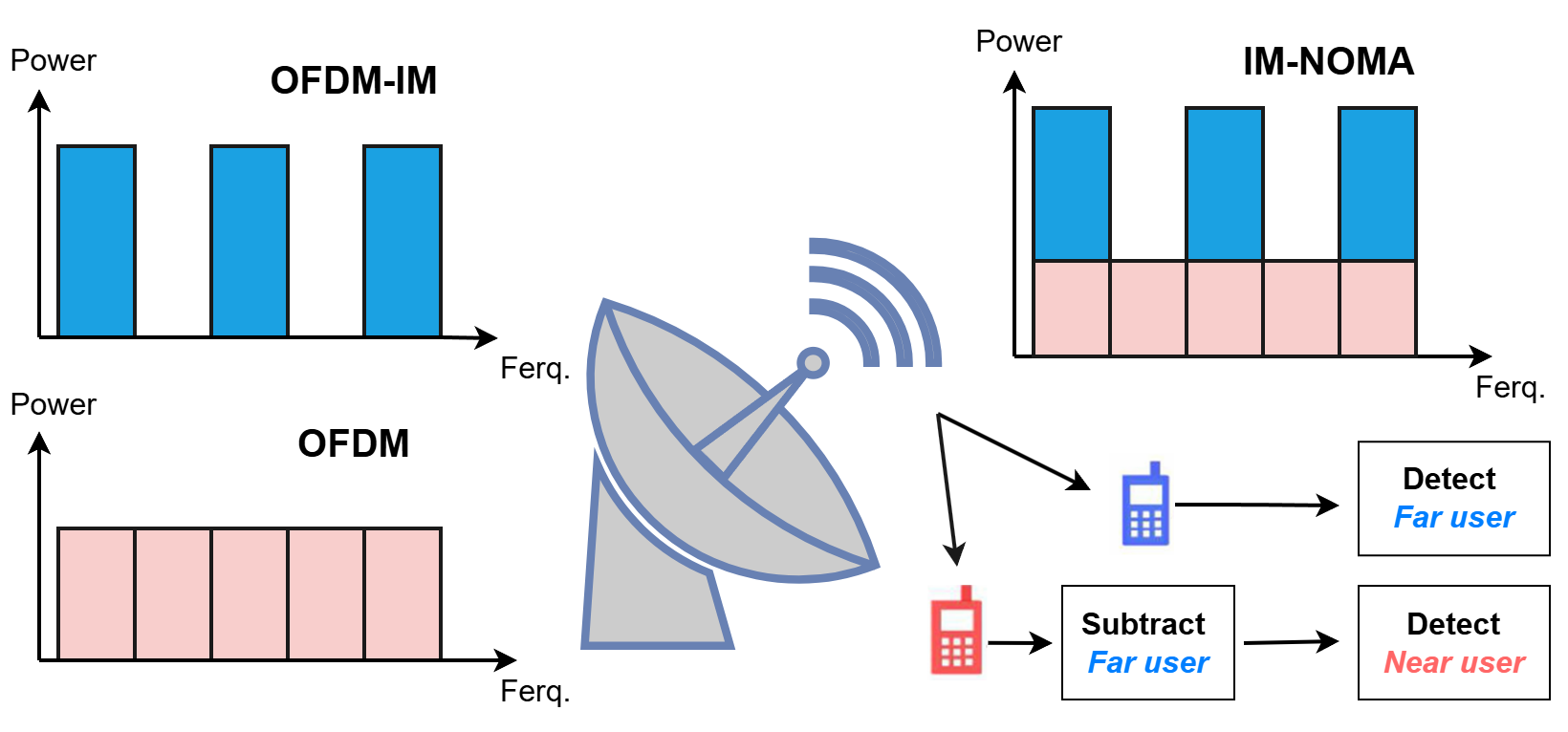} 
\par\end{centering}
\caption{Overview of the hybrid downlink IM-NOMA transmitter.}
\label{fig1}
\end{figure}

\section{Hybrid IM-NOMA system model and baseline detectors
\label{sec:System-Model}}
The hybrid downlink IM-NOMA system exemplifies a cutting-edge approach to maximizing spectral efficiency and user capacity in wireless communications. This system ingeniously combines the strengths of OFDM-IM and NOMA, resulting in a versatile and efficient transmission framework.

\subsection{Hybrid Downlink IM-NOMA System}

In this system, data transmission occurs over multiple blocks, each corresponding to different users or user groups. This layered approach allows simultaneous data transmission across the same frequency spectrum, significantly enhancing spectral efficiency.

\subsubsection{Traditional OFDM Block}
In the conventional OFDM layer, all  \( N \) subcarriers are available for symbol transmission. For user $i$, each subcarrier carries an $M$-ary constellation point, such as a QAM symbol. This full-subcarrier mapping provides a standard and robust baseline for multicarrier transmission, particularly in frequency-selective environments where the cyclic-prefix OFDM structure helps mitigate inter-symbol interference. The baseband OFDM signal of user $i$, $\mathbf{x}_\text{OFDM} \in \mathbb{C}^{1\times N}$, is represented as:
\begin{equation}
\textbf{x}_{\text{OFDM}} = [c_i(1), c_i(2),..., c_i(N)]^T,
\end{equation}
where the $c_i(n)$ represents the constellation symbol transmitted on the $n$-th subcarrier.


\subsubsection{OFDM-IM Block} 
In the OFDM-IM layer, only selected subcarriers are activated. Therefore, the information bits are mapped not only to the symbols and their phases but also to the indices of the active subcarriers. The signal $\mathbf{x}_{\text{OFDM-IM}}\in\mathbb{C}^{1\times N}$ is intended for the far user through $N$ subcarriers, and its entries are defined as:


\begin{equation}
\mathbf{x}_{\text{OFDM-IM}}=\begin{cases}
{c}_{n}, & \text{if}\text{\,\,}n\text{\,\ensuremath{\in \alpha_{i}}}\\
0, & \text{if}\text{\,\,}n\text{\,\ensuremath{\notin \alpha_{i}}}\\
\end{cases}\,\,\,\,\,\forall \,\, n = 1, 2,, ..., N,\label{eq:signal-IMNOMA}
\end{equation}
Here, $\text{c}_{n}$ is the $M$-ary constellation symbol assigned to the $n$-th subcarrier, and $\alpha_{i}$ denotes the active-index set of the OFDM-IM user. The total number of bits in one OFDM-IM symbol includes the modulation bits and the index bits carried by the activated subcarriers:


\begin{equation}
p_{\text{total}} = p_{\text{mod}} + p_{\text{index}} = [ \log_2 \binom{N}{K} ] + K\log_2 M,
\end{equation}
where \(p_{\text{mod}}\) and \(p_{\text{index}}\) denote the bits conveyed by the modulation symbols and the index pattern, respectively. Set \( K \) to be the number of active subcarriers. Among $N$ subcarriers, there are C(N,K) possible active-index choices, and each active tone carries \( \log_2 M \) bits. The active set can be selected either by a combinatorial rule or by a lookup table \cite{OFDM-IM}.


\subsubsection{NOMA Block} 
The NOMA layer multiplexes users in the power domain before transmission. User signals assigned to different power coefficients are superposed and sent over the same frequency resource. In the receiver, each user detects its desired component from the composite signal. In this downlink setting, the far-user stream is allocated higher power because it is more affected by path loss and by the OFDM-IM structure, whereas the near-user stream is recovered after subtracting the reconstructed far-user component. In the frequency domain, the base-station signal is:


\begin{equation}
\textbf{x} = \sum_{i=1}^{L} \sqrt{P_i} \textbf{x}_i,
\end{equation} 
where $\mathbf{X}_i$ is the signal vector of user \(i\), \(P_i\) is the allocated power coefficient of user \(i\), and \(L\) is the number of users. This superposition enables simultaneous resource sharing and can improve spectral efficiency.


The transmission occurs over Rayleigh fading channels, which accurately model multipath propagation effects in wireless environments. The received signal at user \(i\)-th via $N$ subcarriers is:

\begin{equation}
\textbf{y}_i = \textbf{h}_i \textbf{x} + \textbf{n}_i = \textbf{h}_i \textbf{x} = \sum_{i=1}^{L} \sqrt{P_i} \textbf{x}_i + \textbf{n}_i ,
\end{equation} 
here \(\mathbf{h}_i = [h_{i}(1), h_{i}(2), \ldots, h_{i}(N)]^T\) and \(\mathbf{n}_i = [n_{i}(1), n_{i}(2), \ldots, n_{i}(N)]^T\). The vector \(h_i\) contains the complex channel gains of user \(i\), and \(n_i\) denotes additive white Gaussian noise (AWGN) with zero mean and variance \(\sigma^2\). Under this model, the BS broadcasts the composite block to all users, which then recover their target signals using detection methods. Previous research has primarily focused on two key methods: ML and SIC detection. The following sections will explore these methods, along with their associated challenges and limitations.


\subsection{Conventional Signal Detectors}
\subsubsection{Maximum Likelihood Detection (ML)}

ML detection is considered a theoretically optimal method for signal detection, as it seeks to find the best solution in the signal space that minimizes detection errors. For the IM-NOMA signal model, ML selects the solution that minimizes the Euclidean distance between \(y_i\) and all feasible transmitted candidates  \( \textbf{x} \). The decision rule is:


\begin{equation}
\hat{\mathbf{x}}
=
\arg\min_{\mathbf{x}_i \in \delta}
\left\|
\mathbf{y}_i
-
\mathbf{h}_i
\sum_{i=1}^{L}
\sqrt{P_i}\,\mathbf{x}_i
\right\|^{2},
\label{eq:ML_equation}
\end{equation}
where \(\mathbf {h} \) is the channel matrix, and \( \delta \) represents the possible transmitted signals. Although ML is accurate, it becomes computationally expensive as the number of users and subcarriers grows.


\subsubsection{Successive Interference Cancellation (SIC)}
In the IM-NOMA system, users are typically differentiated by their power levels, with users receiving stronger signals having higher power to compensate for greater path loss. SIC offers a more practical alternative to ML, with reduced computational complexity, especially in NOMA systems. Compared with ML, SIC requires less computation, but its performance depends on the separation between user power levels. At the receiver, the highest-power signal is detected first, often corresponding to the user with the weakest channel condition:

%
\begin{equation}
\hat{\textbf{x}}_1 = \textbf{h}_1\mathbf{y} - \sum_{i=2}^{L} \sqrt{P_i} \mathbf{x}_i.
\end{equation} 
After this component is estimated, it is removed from the received signal, and the receiver proceeds to decode the next strongest component:

%
\begin{equation}
\hat{\textbf{x}}_2 = \mathbf{y} - \textbf{h}_1 \hat{\textbf{x}}_1.
\end{equation}

This process repeats until all user signals are recovered. SIC provides a balanced approach between error performance and computational complexity. Yet, it can suffer from error propagation, where errors in detecting the first user can negatively impact the detection of subsequent users.  SIC, with its lower complexity, is more suitable for practical applications, though it may not always achieve the optimal performance seen with ML, especially under challenging channel conditions.

\section{Proposed DeepHSIC detector\label{sec:proposedDeep}}

This section details the architecture and operational principles of the DeepHSIC detector also training procedure. 

\begin{figure}[tb]
\centerline{\includegraphics[width=\textwidth]{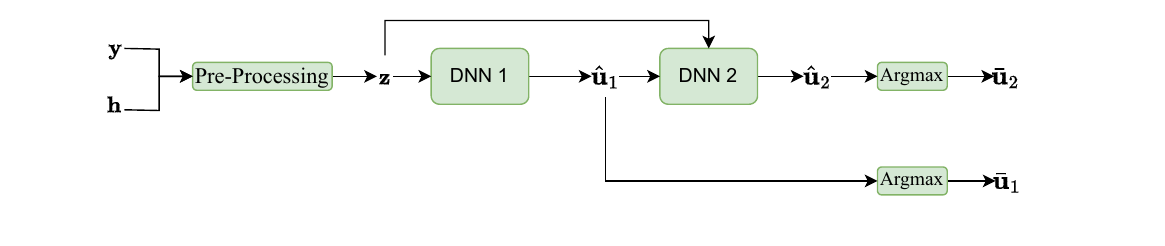}}
\caption{Signal flow of the proposed DeepHSIC detector for user 2 (i.e., $l=2$).\label{fig:fig2}}
\end{figure}

\begin{figure}[tb]
\centerline{\includegraphics[width=\textwidth]{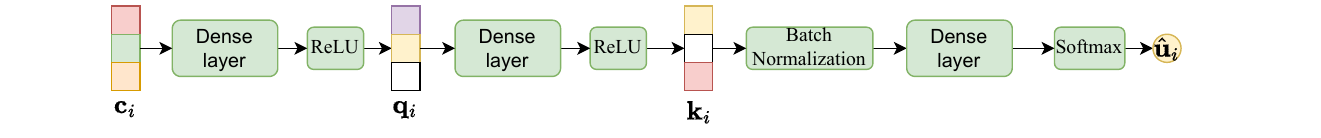}}
\caption{Detailed architecture of the proposed $i$-th DNN of our DeepHSIC.\label{fig:fig3}}
\end{figure}

\subsection{Proposed Network Architecture}
The architecture of the DeepHSIC detector, as illustrated in Fig. \ref{fig:fig2}, includes a pre-processing stage followed by two DNN blocks. For the two-user case, the detector consists of a preprocessing unit followed by two DNN blocks. In the preprocessing unit, the received vector \textbf{y} and the channel-state information \( \textbf{h} \) are converted into a feature vector \( \textbf{z} \). The processing is based on a zero-forcing equalization step, which reduces channel distortion and produces an input that is easier for the neural detector to learn. Specifically, the inverse channel matrix is applied to \textbf{y}, yielding $\tilde{\textbf{y}} = \textbf{h}^{-1}\textbf{y}$, where \( \tilde{\textbf{y}} \) denotes the equalized signal. The real and imaginary parts of \( \tilde{\textbf{y}} \) are then concatenated to form \( \textbf{z} \):


\begin{equation}
\mathbf{z}
=
\bigl[
\operatorname{Re}(\tilde{\mathbf{y}}),
\operatorname{Im}(\tilde{\mathbf{y}})
\bigr],  
\end{equation}
where $\operatorname{Re}(.)$ and $\operatorname{Im}(.)$ are represent for real and imaginary parts. This transformation is crucial as it normalizes and conditions the data, making it suitable for deep learning models. The output \( \textbf{z} \) serves as the input to the first DNN block.

 The core of the DeepHSIC detector consists of two DNN blocks, each responsible for detecting the transmitted symbols for a specific user, enhancing detection accuracy. Initially, the pre-processed signal \( \textbf{z} \) is input into the first DNN block (DNN1), producing an initial estimate \( \hat{\textbf{u}}_1 \) for the first user, described as $\hat{\textbf{u}}_1 = f_1(\textbf{z}; \theta_1)$, where \( f_1 \) is the DNN-based function and \( \theta_1 \) represents its learnable parameters. This estimate, along with the input signal \( \mathbf{z} \), is provided to the second DNN to refine the detection of the next user
, describe as $\hat{\textbf{u}}_2 = f_2(\hat{\textbf{u}}_1, \textbf{z}; \theta_2)$, where \( f_2 \) is the DNN-based function and \( \theta_2 \) represents its learnable parameters. This process mitigates inter-user interference and improves the final decision's accuracy. The outputs from each DNN are processed through an argmax function to yield the detected symbols. 

 \subsection{Detailed DNN Structure}
Each DNN block contains fully connected layers combined with activation and normalization operations, as depicted in Fig.~\ref{fig:fig3}. 
Firstly, the input vector undergoes the FC layer with ReLU activation function, defined as:

\begin{equation}
\textbf{q}_i = \text{ReLU}(\textbf{W}_{1i} \textbf{c}_i + \textbf{b}_{1i}),
\end{equation}

\begin{equation}
\textbf{k}_i = \text{ReLU}(\textbf{W}_{2i} \textbf{q}_i + \textbf{b}_{2i}),
\end{equation}
where $\textbf{W}_{1i}$, $\textbf{W}_{2i}$, $\textbf{b}_{1i}$, and $\textbf{b}_{2i}$ (all known as $\theta_i$) is weight and bias of FC layer $1$ and $2$ of user $i$, respectively, while $\textbf{c}_i$ is the input vector of user $i$, in which  $\textbf{c}_1 = \textbf{z}$ and $\textbf{c}_2 = [\textbf{z};\hat{\textbf{u}}_1]$ The output then undergoes batch normalization, which normalizes the activations over a batch of data, improving training stability and convergence speed. The final layer maps the hidden features to a probability distribution over all feasible transmitted symbols using \text{Softmax}, and the detected symbol is selected as the index with the largest output value.
Finally, the corresponding symbol is determined based on the index of unit element in the one-hot vector $\bar{\textbf{u}}_i$.

\subsection{ Training procedure \label{subsec:3b}}

 The training process begins with the creation of a comprehensive dataset, consisting of randomly generated $b$-bit sequences for $L$ users, encoded into vectors $\textbf{x}_i$ and corresponding one-hot vectors $\textbf{u}_i$, where $i\in L$. The network is trained by minimizing the mean-squared error between the predicted vector $\hat{\textbf{u}}_i$ and the target one-hot vector \( \textbf{u}_i \), defined as:

\begin{equation}
\mathcal{L}(\textbf u_i, \hat{\textbf{u}}_i) = \frac{1}{L} \sum_{i=1}^{L} (\textbf u_i - \hat{\textbf{u}}_i)^2,
\end{equation}
where $\hat{\textbf{u}}_i$ is the output of the DNN model and \(\textbf u_i \) is the ground-truth label. 
To optimize the parameters, The Adam optimizer \cite{adam2014} adjusts the learning rates adaptively to minimize this loss, updating parameters $\theta$ based on the gradients:

\begin{equation}
\theta \leftarrow \theta - \eta \nabla_\theta L(\textbf u_i, \hat{\textbf{u}}_i),
\end{equation} 
where \( \eta \) represents the learning rate. Adam's adaptive learning rates and momentum \cite{adam2014} assist in accelerating convergence and avoiding local minimal. During training, the signal-to-noise ratio (SNR), denoted as \( \lambda_{\text{train}} \), is varied to expose the model to different channel conditions. This variation improves the model’s ability to operate across a range of environments and strengthens its robustness.


\begin{table}
\caption{Simulation setup used in the experiments}
\label{tab:para}
\centering
\begin{tabular}{lr}
\toprule
Parameter  & Value\\
\midrule
Number of users $L$  & 2\\
Parameters of Index Modulation ($N$, $K$, $M$)  & 4, 1, 4\\
Power allocation coefficient ${P}_{1}$--${P}_{2}$  & 0.8--0.2\\
Hidden nodes of DNN 1  & 32--64\\
Hidden nodes of DNN 2  & 128--256\\
Activation function for hidden layers   & ReLU\\
Activation function for output layers & Softmax\\
Training SNR $\lambda_{\text{train}}$  & 35 dB\\
Learning rate $\eta$  & 0.001\\
Batch size  & 1000\\
Number of training epochs  & 100\\
Training data size  & $2\times10^6$\\
Testing data size  & $10^{6}$\\
Optimizer  & Adam \cite{adam2014}\\
\bottomrule
\end{tabular}
\end{table}

\section{Simulation results\label{sec:results}}

This section reports the numerical evaluation of DeepHSIC. The experiments compare the proposed detector with ML and SIC in terms of BER and detection runtime, thereby assessing both reliability and implementation feasibility for the hybrid IM-NOMA system


\subsection{Simulation Parameter Setup}
The simulation parameters for  DeepHSIC are summarized in Table \ref{tab:para}.The system is configured with two users, using index modulation parameters \( (N, K, M) = (4, 1, 4) \), which are defined in Section~\ref{sec:System-Model}. The power allocation coefficients for the users are \( P_1 = 0.8 \) and \( P_2 = 0.2 \). The hidden layers of DNN are configured with $32$-$64$ nodes for DNN 1 and $128$-$256$ nodes for DNN 2. ReLU is used in the hidden layers, and Softmax is used at the output. The training SNR is fixed at 35 dB, the learning rate is \( \eta \) = 0.001, the batch size is 1000, and the model is trained for 100 epochs.


Channel-estimation errors are included to test robustness. The estimated channel is modeled as  \( \hat{h}_l = h_l + e \), where e follows a Gaussian distribution with zero mean and variance \( \sigma_e^2 \). Two cases are considered: perfect CSI with \( \sigma_e^2 = 0 \) and imperfect CSI with \( \sigma_e^2 = 0.01 \). This setting allows the DeepHSIC detector to be evaluated under both ideal and nonideal channel knowledge.


\subsection{BER Performance}

\begin{figure}[h!]
\begin{centering}
\includegraphics[width=0.8\textwidth]{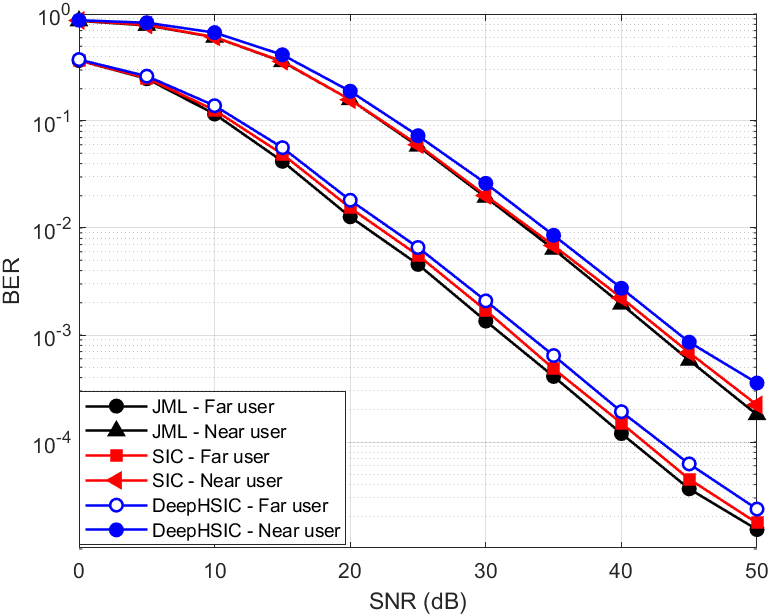} 
\par\end{centering}
\caption{BER results of DeepHSIC and conventional detectors under perfect CSI.\label{fig:user2-perfectCSI}}
\end{figure}

\begin{figure}[tb]
\begin{centering}
\includegraphics[width=0.8\textwidth]{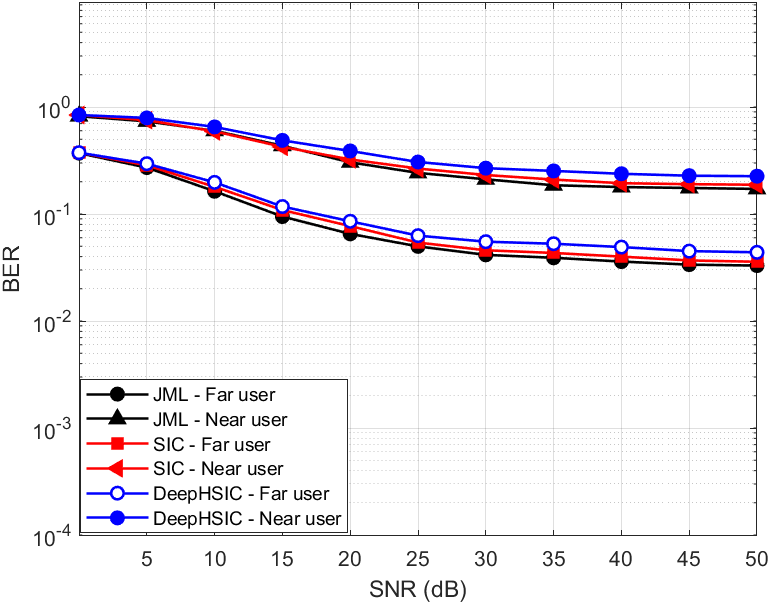} 
\par\end{centering}
\caption{BER results of DeepHSIC and conventional detectors under imperfect CSI. \label{fig:user3-perfectCSI}}
\end{figure} 
We compare DeepHSIC with ML and SIC detectors. First, BER is evaluated under perfect CSI for SNR values from $0$ dB to $50$ dB. As shown in Fig.~\ref{fig:user2-perfectCSI}, DeepHSIC achieves near-optimal BER relative to the model-based baselines, with only a small gap from SNR = SNR = $15$ to high-SNR values.
This deviation may be due to the inappropriate setting of $\lambda_{\text{train}}$ for DNN-based detection under high SNRs. The results indicate that DeepHSIC can learn SIC effectively when trained with data from a single SNR.

In the Fig. \ref{fig:user3-perfectCSI}, the BER of DeepHSIC is compared under imperfect CSI conditions. Despite the degradation due to CSI, DeepHSIC still achieves comparable performance to traditional detectors, leveraging DNNs to decode soft information for each symbol. This emphasises our approach remains effectively in imperfect CSI condition and significantly reduced computational complexity, as discussed in the next section.

\subsection{Runtime Complexity}
Runtime is measured during the testing phase for each detected sample. Table \ref{tab:comrun} shows that DeepHSIC requires substantially less detection time than ML and SIC. 
Specifically, DeepHSIC reduces computational time by up to 7 times compared to ML and 3 times compared to SIC. This demonstrates the computational efficiency of DeepHSIC, making it particularly suitable for real-time applications, where processing speed is critical.

\begin{table}
\centering
\caption{. Per-sample detection runtime of DeepHSIC, ML, and SIC in milliseconds}
\label{tab:comrun}
\begin{tabular}{cccc}
\toprule
$(N,K,M)$ & ML & SIC & DeepHSIC\\
\midrule
(4,1,4) & $7.65\times10^{-4}$ & $3.32\times10^{-4}$ & $1.14\times10^{-4}$\\
\bottomrule
\end{tabular}
\end{table}

\section{Conclusions}
\label{sec:consuc}
We proposed DeepHSIC, a detector for hybrid downlink IM-NOMA systems that replaces the computationally intensive SIC decision block with DNN modules. The detector jointly addresses the OFDM-IM and OFDM signal components without modifying the mathematical system model. Simulation results show near-optimal BER compared with conventional detectors under both perfect and imperfect CSI, while achieving lower runtime. Therefore, DeepHSIC provides a scalable receiver design for modern communication systems and can be extended to larger user configurations and time-varying channels.\\


\section*{Acknowledgement\label{sec:Acknowledgement}}
This research was supported by National Economics University under grant number NEU1-2025.02, and by Vietnam National Foundation for Science and Technology Development (NAFOSTED) under grant number 102.05-2025.57.
%
%
%

\begin{thebibliography}{8}

\bibitem{wang2020}
X. Wang and Y. Liu and D. W. K. Ng and R. Schober.: Exploring the potential of index modulation in wireless networks: Advances and applications. IEEE Communications Surveys \& Tutorials \textbf{22}(4), 2648--2672 (2020). 

\bibitem{IMfor5G}
E. Basar.: Index modulation techniques for {5G} wireless networks. IEEE Communications Magazine \textbf{54}(7), 168--175 (2016). 

\bibitem{OFDM-IM}
E. Basar and U. Aygolu and E. Panayirci and H. V. Poor.: Orthogonal frequency division multiplexing with index modulation. IEEE Transactions on Signal Processing \textbf{61}(22), 5536--5549 (2013). 

\bibitem{M-MAfor5G}
Y. Cai and Z. Qin and F. Cui and G. Y. Li and J. A. McCann.: Modulation and multiple access for {5G} networks. IEEE Communications Surveys and Tutorials \textbf{20}(1), 629--646 (2018). 

\bibitem{Basar2013}
E. Basar and U. Aygolu and E. Panayirci and H. V. Poor.: Orthogonal frequency division multiplexing with index modulation. IEEE Transactions on Signal Processing \textbf{61}(22), 5536--5549 (2013). 

\bibitem{Ding2017}
Z. Ding and X. Lei and G. K. Karagiannidis and R. Schober and J. Yuan and V. K. Bhargava.: A survey on non-orthogonal multiple access for {5G} networks: Research challenges and future trends. IEEE Journal on Selected Areas in Communications \textbf{35}(10), 2181--2195 (2017). 

\bibitem{uplink-IM-NOMA}
M. B. Shahab and S. J. Johnson and M. Shirvanimoghaddam and M. Chafii and E. Basar and M. Dohler.: Index modulation aided uplink {NOMA} for massive machine type communications. IEEE Wireless Communications Letters \textbf{9}(12), 2159--2162 (2020). 

\bibitem{Downlink-IM-NOMA}
A. Almohamad and M. O. Hasna and S. Althunibat and K. Qaraqe.: A novel downlink {IM-NOMA} scheme. IEEE Open Journal of the Communications Society \textbf{2}, 235--244 (2021). 

\bibitem{hybrid}
A. Tusha and S. Dogan and H. Arslan.: A hybrid downlink {NOMA} with {OFDM} and {OFDM-IM} for beyond {5G} wireless networks. IEEE Signal Processing Letters \textbf{27}, 491--495 (2020). 

\bibitem{Basar2016}
E. Basar.: Index modulation techniques for {5G} wireless networks. IEEE Communications Magazine \textbf{54}(7), 168--175 (2016). 

\bibitem{Mao2018}
Y. Mao and J. Wu.: A novel NOMA-OFDM with index modulation for {5G} ultra-dense networks. IEEE Communications Letters \textbf{22}(3), 598--601 (2018)

\bibitem{Toan2022APSIPA}
T. Gian and V.-D. Ngo and T.-H. Nguyen and T. T. Nguyen and T. V. Luong.: Deep neural network-based detector for single-carrier index modulation {NOMA}. In: 2022 Asia-Pacific Signal and Information Processing Association Annual Summit and Conference (APSIPA ASC), pp. 1805--1809 (2022). 

\bibitem{SICnet}
T. Van Luong and N. Shlezinger and C. Xu and T. M. Hoang and Y. C. Eldar and L. Hanzo.: Deep learning based successive interference cancellation for the non-orthogonal downlink. IEEE Transactions on Vehicular Technology \textbf{71}(11), 11876--11888 (2022). 

\bibitem{Toan2023}
T. Gian and T.-H. Nguyen and T. T. Nguyen and V.-C. Pham and T. V. Luong.: Transformer-based deep learning detector for dual-mode index modulation {3D-OFDM}. In: 2023 Asia Pacific Signal and Information Processing Association Annual Summit and Conference (APSIPA ASC), pp. 291--296 (2023). 

\bibitem{Thien2019}
T. V. Luong and Y. Ko and N. A. Vien and D. H. N. Nguyen and M. Matthaiou.: Deep learning-based detector for OFDM-IM. IEEE Wireless Communications Letters \textbf{8}(4), 1159--1162 (2019). 

\bibitem{van2022deep}
T. Van Luong and N. Shlezinger and C. Xu and T. M. Hoang and Y. C. Eldar and L. Hanzo.: Deep learning based successive interference cancellation for the non-orthogonal downlink. IEEE Transactions on Vehicular Technology \textbf{71}(11), 11876--11888 (2022). 

\bibitem{adam2014}
D. P. Kingma and J. Ba.: Adam: A method for stochastic optimization. In: ICLR (2014)

\end{thebibliography}
%

\end{document}